\documentclass[12pt]{article}
\pdfoutput=1
\usepackage{color}
\usepackage{amssymb}
\usepackage{amsmath}
\usepackage{graphicx}
\usepackage{amsfonts}
\usepackage{hyperref}

\newcommand{\pprime}{{\prime\prime}}
\newcommand{\lapprox}{\stackrel{_<}{_\sim}}

\DeclareRobustCommand{\SkipTocEntry}[4]{}

\begin{document}

\begin{titlepage}

\setcounter{page}{1} \baselineskip=19.5pt \thispagestyle{empty}

\begin{flushright}
  DESY 12-067
\end{flushright}
\vfil

\begin{center}
{\LARGE A Toy Model For Single Field Open Inflation}

\end{center}
\bigskip\

\begin{center}
{\large Pascal M. Vaudrevange and Alexander Westphal}
\end{center}

\begin{center}
\textit{Deutsches Elektronen-Synchrotron DESY, Theory Group, D-22603 Hamburg, Germany}
\end{center} \vfil

\noindent Inflation in an open universe produced by Coleman-De Luccia
(CDL) tunneling induces a friction term that is strong enough to allow
for successful small-field inflation in models that would otherwise
suffer from a severe overshoot problem. In this paper, we present a
polynomial scalar potential which allows for a full analysis. This
provides a simple model of single-field open inflation on a
small-field inflection point after tunneling. We present numerical
results and compare them with analytic approximations.

\vfil
\begin{flushleft}
%\today
May 8, 2012
\end{flushleft}

\end{titlepage}

\newpage
%\tableofcontents
\newpage

\section{Introduction}
Recently, inflation in open universes has gained significant interest
in the context of the string theory landscape \cite{Susskind:2003kw,
  Sugimura:2011tk, Yamauchi:2011qq,Kleban:2012ph,Guth:2012ww}. Transitions between its plethora
of different vacua most likely happen via tunneling processes
\cite{Coleman:1977py, Coleman:1980aw, Hawking:1981fz}. The resulting
bubble looks like an open Friedmann-Robertson-Walker universe for
observers in the interior. If we assume that at early times, the
universe was trapped in a false vacuum at high energy, it is
conceivable that we are now sitting inside a bubble of (hopefully)
true vacuum. Even though cosmological observations give rather
stringent limits on the curvature of the universe,
$-0.0133<\Omega_k<0.0084$ at $95\%$CL, having an open universe is not
excluded \cite{Komatsu:2010fb}. Quite contrary to this, it appears to
us that there is a strong theoretical prior for $\Omega_k\lapprox0$ if
one accepts the landscape paradigm of string theory. There simply is
no other way known to traverse the vacuum energy landscape but by
tunneling, leading to an open, inflating universe inside the low
energy bubble. In this case, the curvature of the universe will be
redshifted to (probably) unobservably small values by exponential
expansion.

It is quite easy to devise potentials with successful large-field
inflation -- that is, inflation lasts long enough to solve the
well-known cosmological homogeneity, flatness and isotropy
problem. However, these models typically involve field motion over a
distance of several Planck masses. This necessitates protection
  of the potential by a spontaneously broken shift symmetry.
Such a shift symmetry can be derived from a fundamental description such as
string theory, using constructions such as N-flation or axion
monodromy. 

In contrast to this, small-field inflection point inflation models do
not need a symmetry. Instead, they derive their slow-roll flatness
from fine-tuning dimension-6 operators which may be computable in a
given string theory embedding (for a recent review on large-field and
small-field inflation models from string theory, see
e.g.~\cite{Baumann:2009ni}). Small-field inflation models
potentially suffer from a severe overshoot problem~\cite{Brustein:1992nk}.

Open inflation after CDL tunneling was discussed before in the context
of single-field chaotic inflation in a quadratic
potential~\cite{Linde:1998iw}. In this model, the barrier and the
false vacuum were supplied by adding a non-polynomial term to the
standard $m^2\phi^2$ term that has the form of a Lorentz-resonance
line. Here, the negative curvature inside the CDL bubble is not needed
to prevent an overshoot problem: the quadratic potential after exit
from the barrier generates large-field inflation which has a dynamical
slow-roll attractor with a wide field
range. In~\cite{Sugimura:2011tk}, a two-field chaotic open
inflation model was constructed. Here, tunneling proceeds from a high-mass
high-vacuum energy de Sitter minimum predominantly in the direction of
a 2nd non-inflaton scalar field, and exits into a shallow quadratic
large-field inflation potential valley.

In \cite{Dutta:2011fe}, it was shown that in inflation after
tunneling, the resulting friction term -- coming from the curvature
contribution -- in the evolution equation for the inflaton can
alleviate this problem.\footnote{The slowing effect of the negative
  curvature inside a CDL bubble on a scalar field was discussed before
  for a piecewise linear potential in~\cite{Freivogel:2005vv}, and
  noted in general already in~\cite{Dvali:2003vv}.} For monomial
potentials of sufficiently high order, the negative spatial curvature
inside the CDL bubble is strong enough to allow for successful
inflation in small-field models that would otherwise suffer from a
severe overshoot problem. In this paper, we analyze a simple
polynomial single-field scalar potential, encompassing tunnelling from
the false to the inflationary vacuum, as well as a subsequent period
of open inflation.

In Section~\ref{sec:tunneling}, we give a brief review of how to
compute the tunneling process, based on pioneering work by
\cite{Coleman:1977py}. In Section~\ref{sec:open_inflation} we review
inflation in an open universe. Section~\ref{sec:toy_model} contains
the description of tunneling and inflation in a toy model, where we
present numerical results and compare them with analytic
approximations developed in \cite{Dutta:2011ej, Dutta:2011rc}. Finally
we conclude in Section~\ref{sec:conclusions}.

\section{Tunneling}\label{sec:tunneling}
The analysis of tunneling from false to true vacuum in scalar field
was pioneered by \cite{Coleman:1977py, Coleman:1980aw}. The
probability per unit volume for the transition from the false vacuum
at $\phi_+$ to the true vacuum located at $\phi_-$ in a potential
$V(\phi)$ which has a barrier at $\phi_+<\phi_T<\phi_-$ is given by
\begin{eqnarray}
  \frac{\Gamma}{V}&=&A e^{-B}\,,
\end{eqnarray}
where $B$ is given by
\begin{eqnarray}
  B&=&2\pi^2\int\!dr\, r^3\left(\frac{1}{2}\phi_{B}^{\prime 2}+V(\phi)\right)-2\pi^2\int\!dr\,r^3V(\phi_+)\,,
\end{eqnarray}
$r$ the Euclidean radius, and $\phi_B$ is the $O(4)$ symmetric,
so-called bounce solution to the Euclidean equation of motion
\begin{eqnarray}
  \phi_B^\pprime(r)+\frac{3}{r}\phi_B^\prime(r)-\partial_\phi V(\phi)\Big|_{\phi=\phi_B}&=&0\,.
\end{eqnarray}
We work in units $M_p^2=\frac{1}{8\pi G}=1$. In general, the field
$\phi$ does not materialize exactly in the true vacuum at $\phi_-$,
but rather some distance away from it\footnote{Only in the thin-wall
  limit does the field materialize infinitesimally close to the true
  vacuum.} at a point that we denote by $\phi_0$, see
Figure~\ref{fig:V8_no_extra_tilt}(c). In the center of the nucleated
bubble $r=0$, the field is located at $\phi(0)=\phi_0>\phi_T$ with
$\phi^\prime(0)=0$. Crossing the bubble wall located approximately at
$R_T$ (with $\phi(R_T)=\phi_T$), the bounce interpolates between
$\phi_0$ and the false vacuum at $\phi(R_+)=\phi_+$\footnote{Strictly
  speaking, $R_+\to \infty$.}.

\section{Open Inflation}\label{sec:open_inflation}
In this section, we give a brief overview of open inflation driven by
a scalar field. Scalar field dynamics in an open FRW background
\begin{eqnarray}
  ds^2&=&dt^2-a(t)^2\left(\frac{1}{1+r^2}dr^2+r^2d\Omega\right)\,,
\end{eqnarray}
where $a(t)$ is the scale factor, is governed by the following
equations of motion
\begin{eqnarray}
  \ddot{\phi}&=&-3\frac{\dot{a}}{a}\dot{\phi}-\partial_\phi V(\phi)\,,\label{eq:eom:phi}\\
  \frac{\ddot{a}}{a}&=&\frac{1}{3}\left(-\dot{\phi}^2+V(\phi)\right)\,,\label{eq:eom:a}\\
  \left(\frac{\dot{a}}{a}\right)^2&=&\frac{1}{3}\left(\frac{1}{2}\dot{\phi}^2+V(\phi)\right)+\frac{1}{a^2}\,.\label{eq:eom:Friedmann}
\end{eqnarray}
At early times, the friction term in \eqref{eq:eom:phi} is dominated
by the curvature contribution in \eqref{eq:eom:Friedmann}
$\frac{1}{a^2}$, while $a\propto t$. Inflation is defined as
accelerating expansion
\begin{eqnarray}
  \frac{\ddot{a}}{a}&\equiv&\dot{H}+H^2=H^2\left(1-\epsilon\right)>0\,,\label{eq:def:epsilon}
\end{eqnarray}
where the Hubble parameter $H=\frac{\dot{a}}{a}$. We define
$\epsilon=-\frac{\dot{H}}{H^2}$ with $0<\epsilon<1$ for inflation to
take place. Hence we see that for a scalar field in an open FRW
universe with $a\propto t$ at early times, $\epsilon\lapprox1$ due to
the contribution of the potential energy to $H$ in the denominator.

To successfully solve the cosmological flatness and horizon problem,
inflation should last sufficiently long, for approximately $N=60$
efolds or so\footnote{The exact number of required efolds depends on
  the energy scale of both inflation and reheating.}, where the number
of efolds is $N\equiv \ln \frac{a_0}{a}$, $a_0$ is the scale factor at
the end of inflation, and $N$ is measured backwards from the end of
inflation. In other words, $N=0$ when $\epsilon=1$, and $N>0$ for
earlier times.

\section{Toy Model}\label{sec:toy_model}
\begin{figure}
  \begin{center}
  \includegraphics[width=0.8\textwidth]{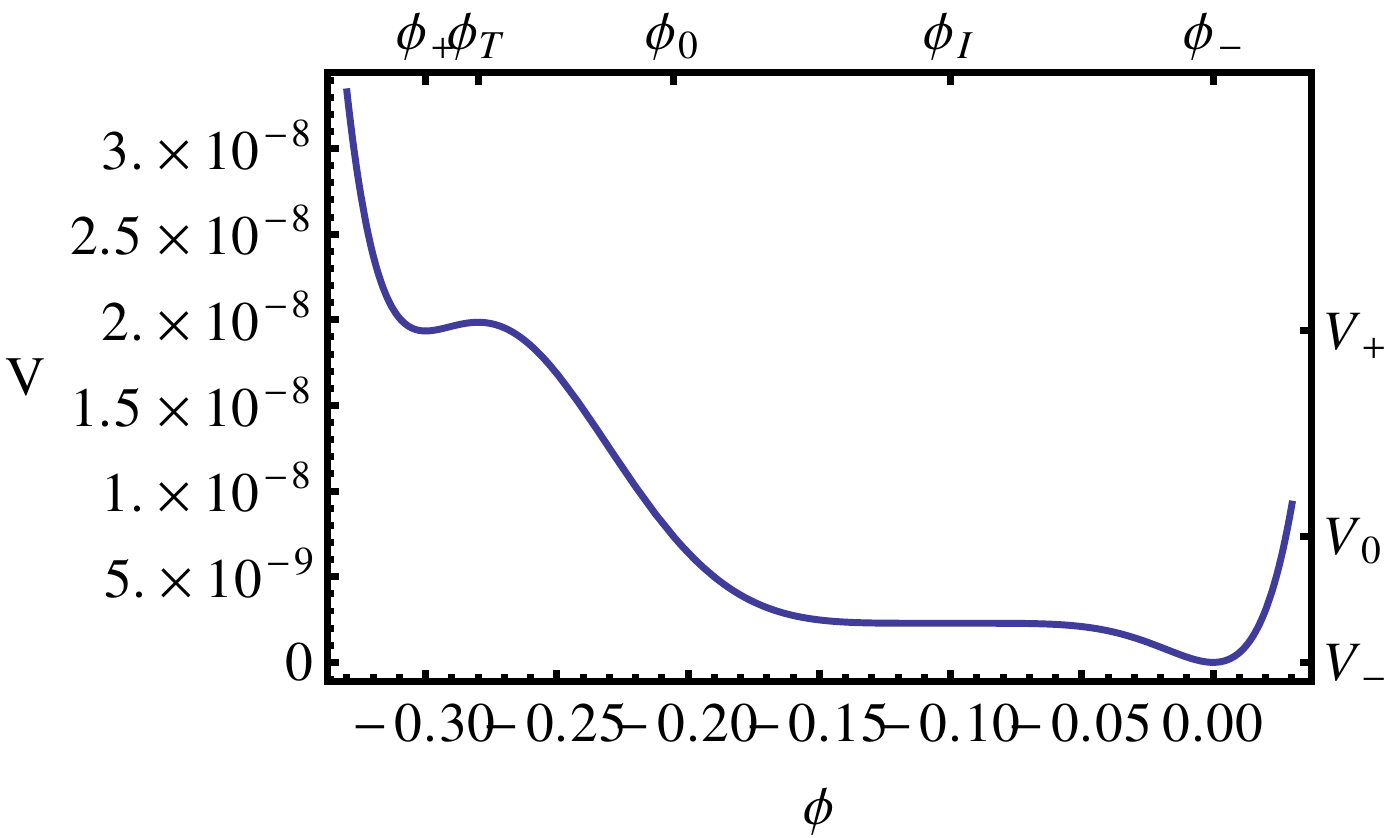}
  \end{center}
  \caption{Plot of the potential. $\phi_+=-\frac{30}{100}$ is the
    location of the false vacuum. $\phi_T=-\frac{28}{100}$ the
    location of the top of the barrier. $\phi_0=-0.205$ is the
    location the field tunnels to. $\phi_I=-\frac{10}{100}$ is the
    inflection point.}
  \label{fig:V8}
\end{figure}
We arrange for the potential to have an inflection point at
$\phi=-\frac{1}{10}$ and further critical points at $-\frac{28}{100},
-\frac{30}{100}, 0$ by writing down the derivative of V
\begin{eqnarray}
  V^\prime(\phi)&=&\phi(\phi+\frac{10}{100})^4(\phi+\frac{30}{100})(\phi+\frac{28}{100})\,,
\end{eqnarray}
and integrating the potential subject to the condition $V(0)=0$. This
gives
\begin{eqnarray}\label{eq:V8_no_extra_tilt}
  V(\phi)&=&\sum_{i=2}^8c_i\phi^i \qquad {\rm with:}\;\; c_2=42\times 10^{-7}\;,\;c_3=\frac{3940}{3}\times 10^{-7}\;,\;c_4=1865\times 10^{-6}\;,\nonumber\\
&&\phantom{\sum_{i=2}^8c_i\phi^i \qquad {\rm with:}}\;\;c_5=1448\times 10^{-5}\;,c_6=\frac{188}{3}\times 10^{-3}\;,\;c_7=14\times 10^{-2}\;,\nonumber\\
&&\phantom{\sum_{i=2}^8c_i\phi^i \qquad {\rm with:}}\;\;c_8=1/8\;,
\end{eqnarray}
see Figure~\ref{fig:V8}. Note that such an 8th order
polynomial potential can be generated quite naturally, e.g.~by
higher-dimension operators present in explicit string theory
constructions of warped D3-brane inflation~\cite{Agarwal:2011wm}. The
false minimum is located at $\phi_+=-\frac{3}{10}$, the potential
barrier at $\phi_T=-\frac{28}{100}$, the inflection point at
$\phi_I=-\frac{1}{10}$ and the true minimum at $\phi_-=0$. The field
tunnels from $\phi_+$ to some position $\phi_0$ located between
$\phi_T$ and $\phi_I$. Then it rolls down the slope towards $\phi_-$,
driving inflation.

\begin{figure*}[t!]
  (a)\includegraphics[width=0.45\textwidth]{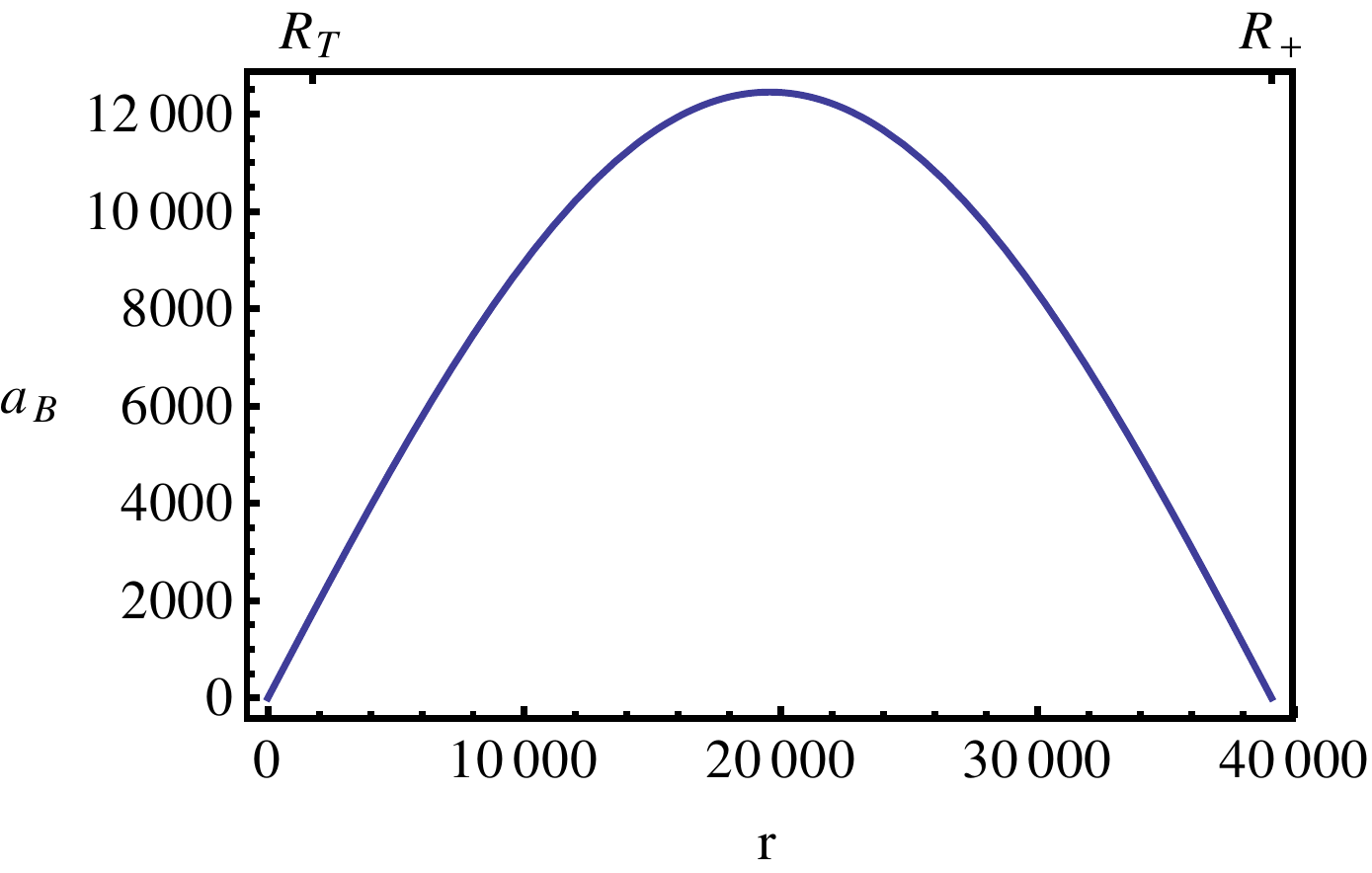}
  (b)\includegraphics[width=0.45\textwidth]{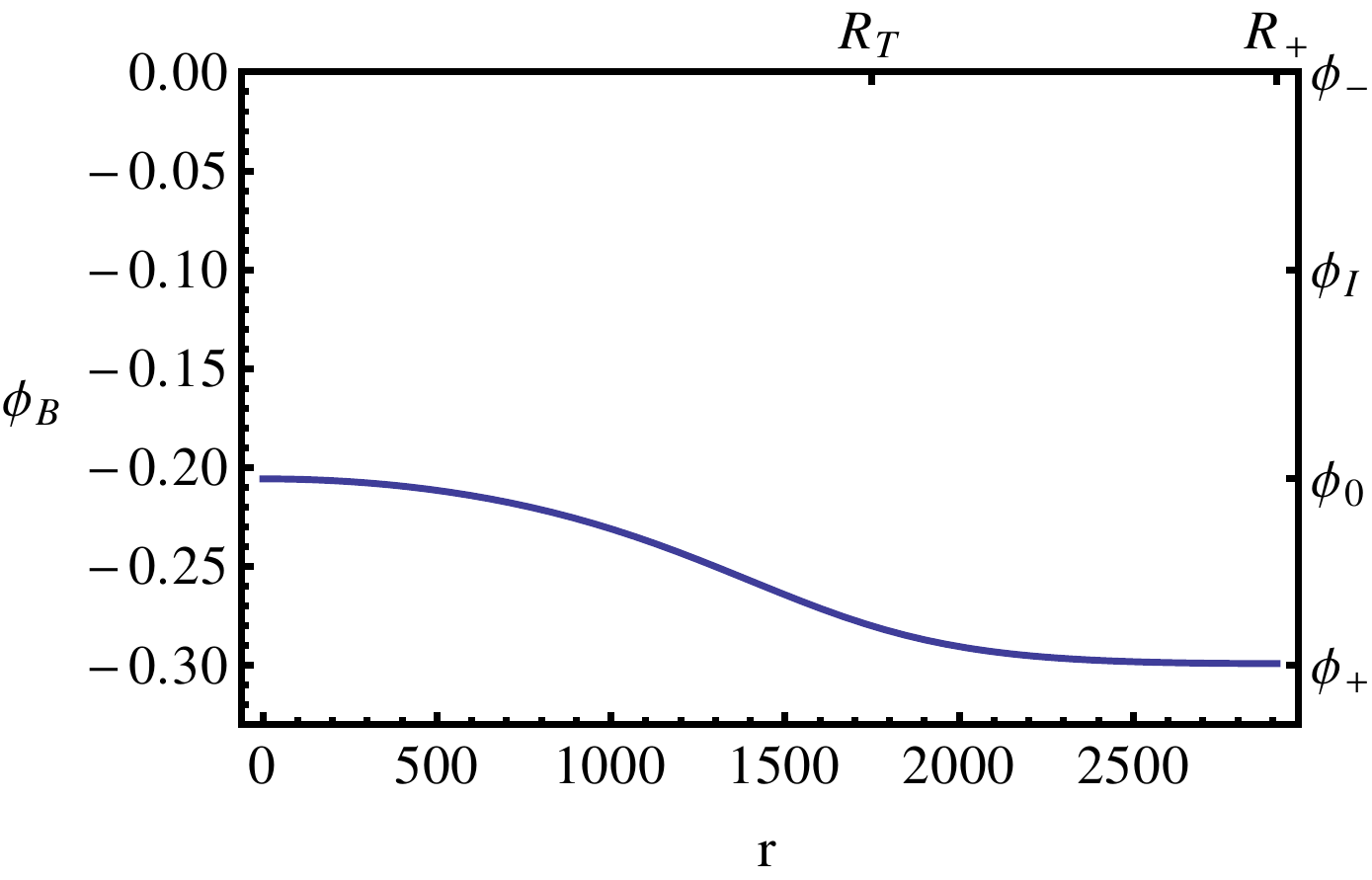}\\
  (c)\includegraphics[width=0.45\textwidth]{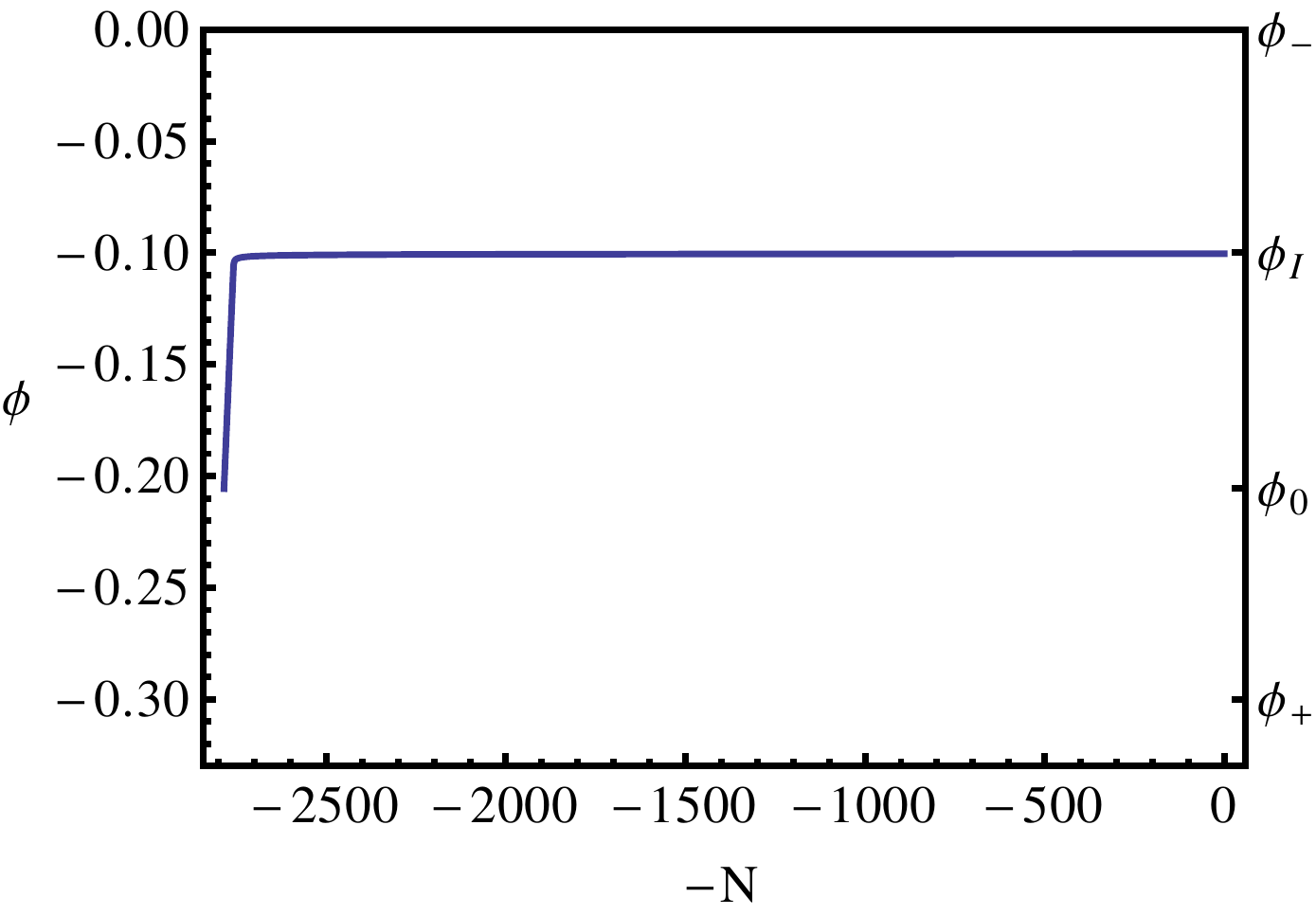}
  (d)\includegraphics[width=0.45\textwidth]{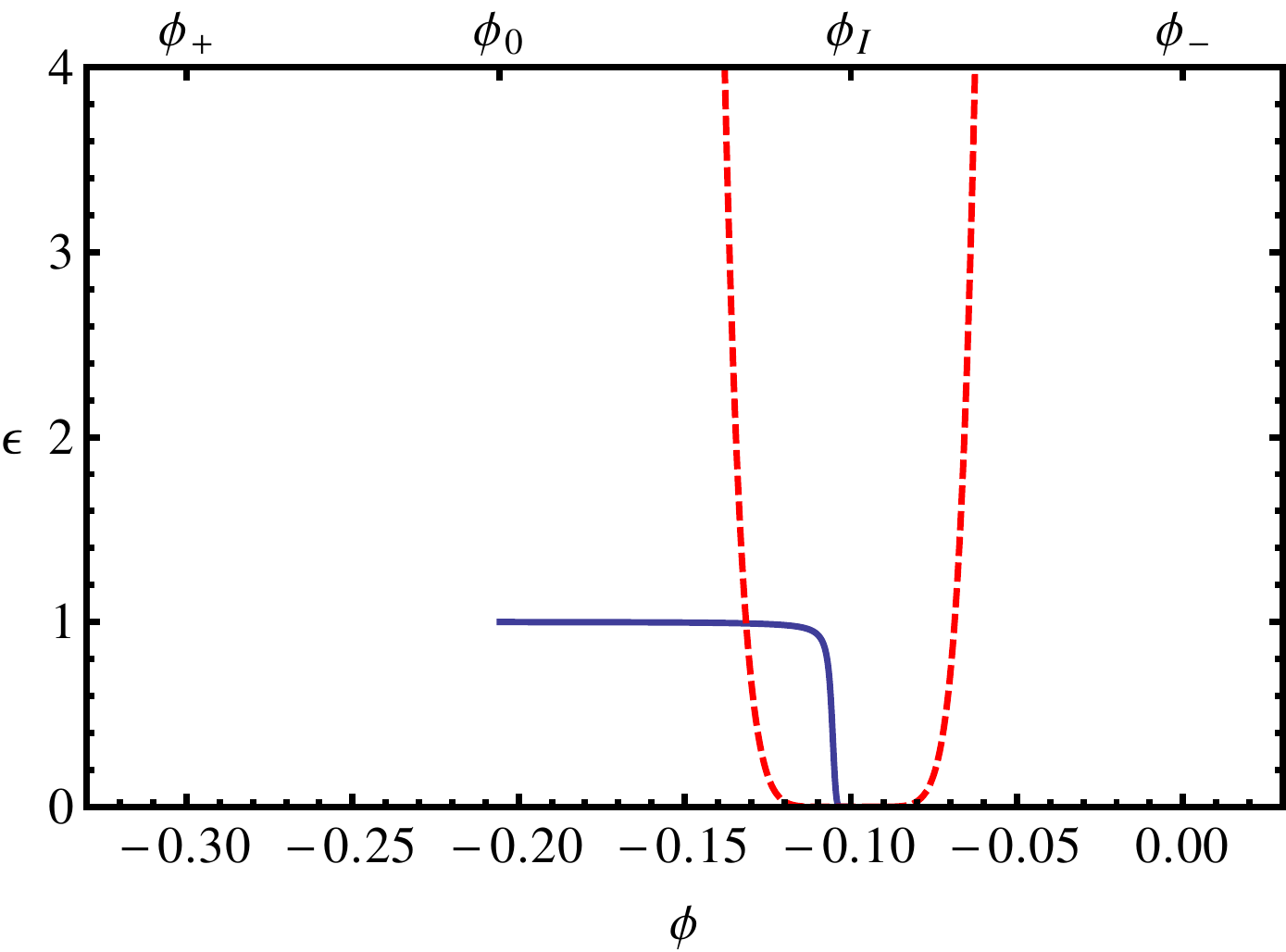}
  \caption{For the potential in \eqref{eq:V8_no_extra_tilt}: (a) Scale
    factor $a$ for the bounce solution. (b) Scalar field $\phi$ for
    the bounce solution, interpolating between $\phi_0$ and
    $\phi_+$. (c) Inflaton trajectory. The field rolls down from
    $\phi_0$ and its classical motion gets stuck at $\phi_I$, so we are just displaying the first $\approx 2500$ of infinitely many e-folds of pure classical slow-roll. The field undergoes slow-roll eternal inflation close to the inflection point. (d) $\epsilon$ as a function
    of the inflaton $\phi$, both the strict definition
    $\epsilon=\epsilon_H=-\dot{H}/{H^2}$ (solid blue) as well as in
    terms of the scalar potential $\epsilon_V=\frac12
    (V'(\phi)/V(\phi))^2$ (dashed red). While $\epsilon_V$ has two
    crossings of $1$, indicating a possible end of inflation, this
    estimate does not correctly capture the real dynamics. As can be
    seen in panel (c), the classical motion of the inflaton comes to a complete halt at
    $\phi_I$. For an analytical explanation, see the main text. Note, that diffusion due to quantum fluctuations of the inflaton in dS space will have the field crossing $\phi_I$ in finite time eventually, and thus lead to a finite maximum of e-folds observed at every given comoving point~\cite{Linde:2007jn}.}
  \label{fig:V8_no_extra_tilt}
\end{figure*}

\subsection{Tunneling process}
While tunneling can in principle proceed either via Coleman de-Luccia
(CdL) \cite{Coleman:1980aw} or Hawking-Moss tunneling
\cite{Hawking:1981fz}, a quick check reveals that due to
$|V^\pprime(\phi_T)|/H^2(\phi_T)\approx900$, the tunneling process is
described by the CdL bounce. This solution to the tunneling problem
can be easily found numerically, see
Figure~\ref{fig:V8_no_extra_tilt}(c). In Euclidean space, the bounce of the full gravitational CdL instanton
interpolates between a point extremely close to the false vacuum $\phi_+=-\frac{3}{10}$ far
outside of the bubble $r\ge R_+$ and $\phi_0\approx -0.2057$ for $r=0$ inside
of the bubble. The Euclidean action for the CdL instanton gives $$\Gamma=e^{-B}=e^{-S_E(\phi_B)+S_E(\phi_+)}\approx e^{-7\times 10^4}\quad.$$

We can apply the results of \cite{Dutta:2011rc} to analytically
estimate the tunneling location $\phi_0$ without gravity. For this purpose, we note in
jumping ahead, that close to the inflection point $\phi_I$ we have
$V(\phi)\sim (\phi-\phi_I)^5$. However, an analytic expression
  for the bounce in a quintic potential is not known. Therefore we
choose to approximate the potential by a piecewise quartic-quartic
potential with the same values for $\phi_+,\phi_T,\phi_I$ and
$V_+,V_T,V_I$ as used for the original scalar
potential~\eqref{eq:V8_no_extra_tilt}.  Equations~(6) and (8) in
\cite{Dutta:2011rc} give the value of $\phi_0$ for tunneling in such a
piecewise quartic-quartic potential excluding the effects of gravity
as
\begin{eqnarray}
  \phi_0&=&\phi_T+\left(\frac{1-\sqrt{1+z^2}}{z}\right)^2(\tilde{\phi}_--\phi_T)\approx-0.26\,,
\end{eqnarray}
where
\begin{eqnarray}
  z=\frac{(1+2\alpha)\sqrt{\Delta}+\sqrt{4\alpha(1+\alpha)+\Delta}}{1-\Delta}\,,\quad
  \Delta=\frac{V(\phi_T)-V(\phi_+)}{V(\phi_T)-V(\tilde{\phi}_-)}\,,\quad
  \alpha=\frac{\phi_+-\phi_T}{\phi_T-\tilde{\phi}_-}\,.
\end{eqnarray}
The additional terms $\phi_T$ appear because in \cite{Dutta:2011rc},
the coordinate system was chosen in such a way as to have
$\phi_T=0$. While the true minimum of the
potential~\eqref{eq:V8_no_extra_tilt} is located at $\phi_-$, it is
clear that the more appropriate ``true'' minimum to use is
$\tilde{\phi}_-=\phi_I$. Hence we obtain $\phi_0\approx-0.26$ which is
in agreement with the numerical results.

\subsection{Inflationary trajectory}
Numerically solving the field dynamics, Equations \eqref{eq:eom:phi}
and \eqref{eq:eom:a}, for the inflationary period, we find that the
field stops at the inflection point. Consequently, inflation never
ends, see Figure~\ref{fig:V8_no_extra_tilt}(c), which shows the
evolution of $\epsilon$ \eqref{eq:def:epsilon} as a function of the
number of efolds. Initially, $\epsilon$ is only a little smaller than
unity. As outlined in Section~\ref{sec:open_inflation}, this can be
understood as a consequence of the fact that for open inflation,
$a\propto t$ at early times. In fact, the friction term in
Equation~\eqref{eq:eom:phi} dominates the potential contribution and
causes a slow down of the scalar field, see
Figure~\ref{fig:V8_no_extra_tilt}(d). Together with the constant
contribution to $H$ from the potential plateau, this is responsible
for the complete stop of the scalar field at the inflection
point. Ignoring the curvature term, the inflaton overshoots the
inflationary plateau within less than an efold. This kind of behavior
was outlined in \cite{Dutta:2011fe}. There, we found that for open
inflation in monomial potentials of order $\ge4$ without uplifts, the
inflaton comes to a stop at the inflection point. Writing the
potential \eqref{eq:V8_no_extra_tilt} as expansion around the
inflection point $\phi_I=-\frac{1}{10}$, we find\footnote{Note that
  this expression is exact as we are dealing with a finite power
  series.}
\begin{eqnarray}
  V(\phi_I+\delta\phi)&=&0.023-7200\,\delta\phi^5+{\cal O}(\delta\phi^6)\,,
\end{eqnarray}
which shows that the potential is sufficiently flat around $\phi_I$
for the results of \cite{Dutta:2011fe} to apply. Strictly speaking,
the analysis was performed for potentials with $V(\phi_I)=0$ and
additionally assuming curvature domination throughout. While the
eventual failure of curvature domination leads to a non-zero velocity
at $\phi_I$. However, the presence of a non-zero uplift $V(\phi_I)>0$
more than compensates for this such that the induced friction term can
easily make the inflaton come to a complete stop.

\subsection{Tilted model}
\begin{figure*}[ht]
  (a)\includegraphics[width=0.45\textwidth]{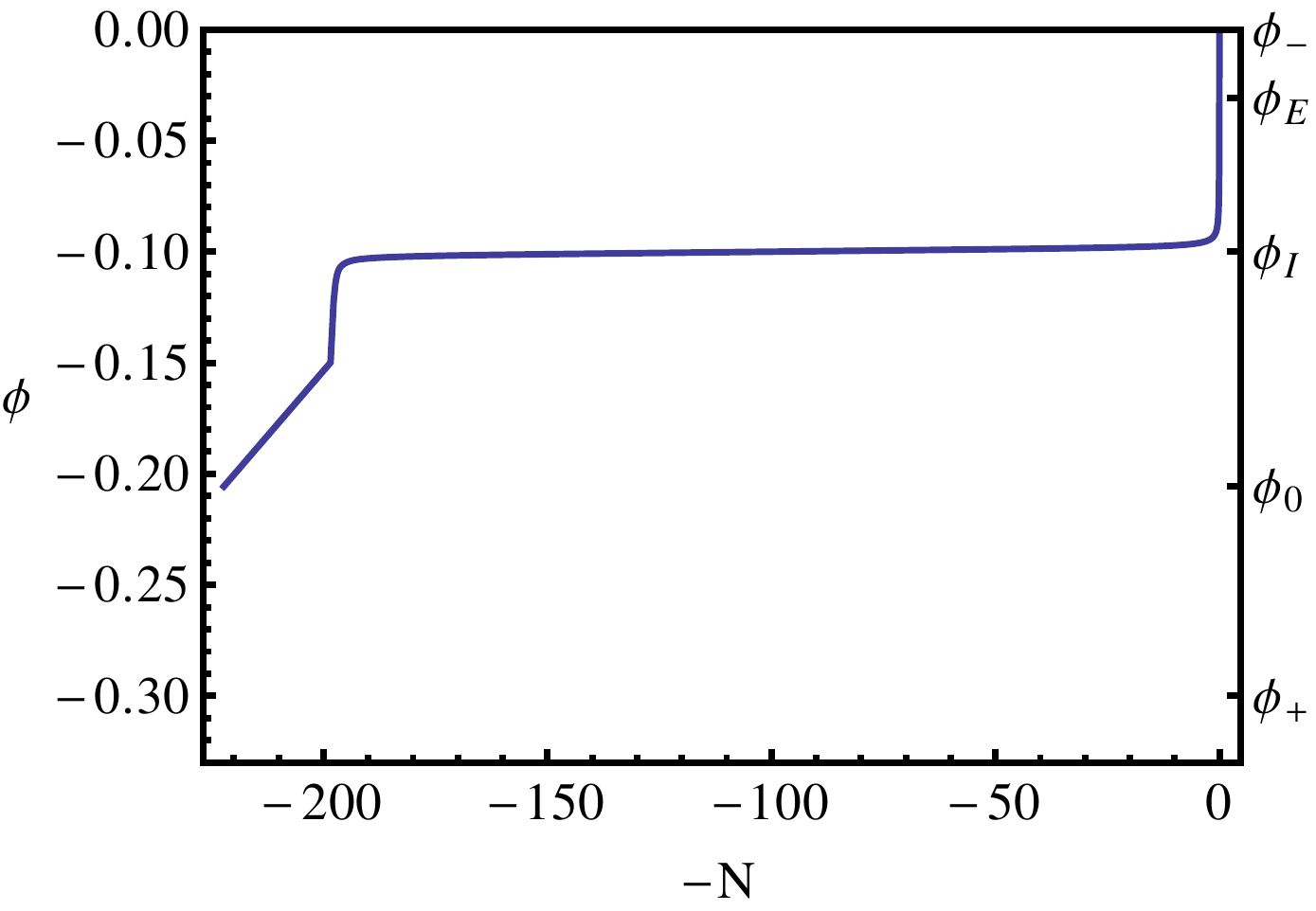}\
  (b)\includegraphics[width=0.45\textwidth]{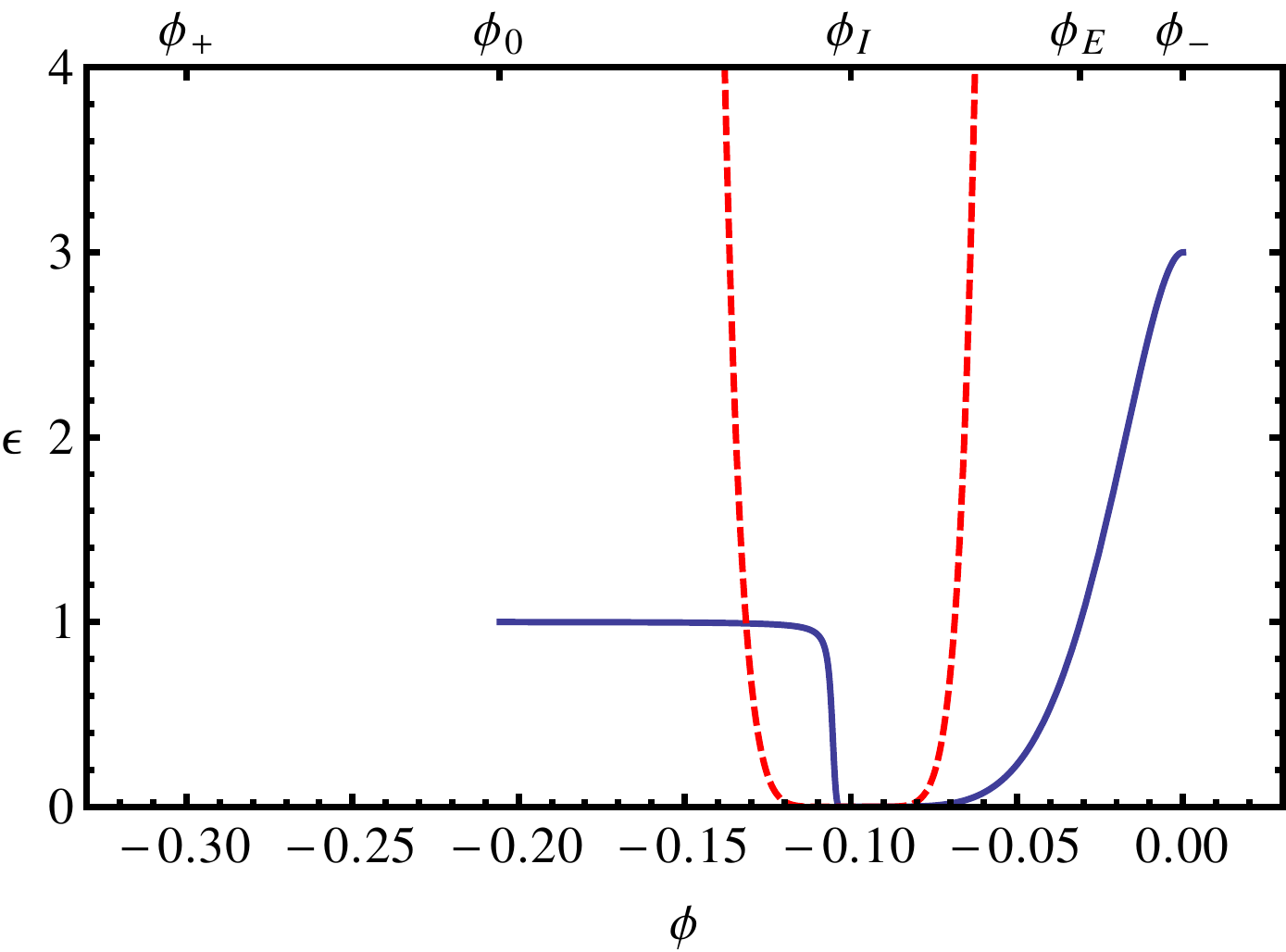}
  \caption{For the potential in (\ref{eq:V8_extra_tilt}): (a) The
    inflationary trajectory in the inflationary universe on the inside
    of the bubble. While the field rolls down from $\phi_0$,
    $\epsilon$ is just a little smaller than $1$, i.e. inflation is
    taking place. Then, the field enters a slow roll phase around the
    inflection point $\phi_I$ (without slow-roll eternal inflation) due to the combined effect of the
    curvature and the potential energy $V(\phi_I)>0$. Finally, the
    field rolls off and accelerated expansion stops at $\phi_{E}$. (b)
    $\epsilon$ as a function of the number of efolds, both the strict
    definition $\epsilon=\epsilon_H=-\dot{H}/{H^2}$ (solid blue) as
    well as in terms of the scalar potential $\epsilon_V=\frac12
    (V'(\phi)/V(\phi))^2$ (dashed red). At the end of inflation at
    $\phi_{E}$, we have $\epsilon_H=1$.}
  \label{fig:V8_extra_tilt}
\end{figure*}

As we strive to develop a model where inflation eventually ends, we
simply add a small negative slope and a corresponding uplift (to
ensure $V(\phi_-)=0$) to the potential
\begin{eqnarray}\label{eq:V8_extra_tilt}
  V(\phi)&=&1.5\times10^{-15}-5\times10^{-7}\,\phi+\sum_{i=2}^8c_i\phi^i \,,
\end{eqnarray}
resulting in a finite duration of inflation, see
Figure~\ref{fig:V8_extra_tilt}.
Introducing the extra tilt does not change position of $\phi_+,
\phi_T, \phi_0, \phi_I, \phi_-$ by much. It does however alter the
expansion of the potential around the inflection point
\begin{eqnarray}\label{Vapprox}
  V(\phi_I+\delta\phi)&=&0.023-5\times10^{-7}\,\delta\phi-7200\,\delta\phi^5+{\cal O}(\delta\phi^6)\,,
\end{eqnarray}
introducing a linear term there as well which will be responsible for
the end of inflation. As pointed out in \cite{Dutta:2011fe}, the field
can enter slow-roll already to the left of the inflection point at
$\phi_I$ due to vacuum energy domination from the substantial uplift
$V(\phi_I)=0.023$. Writing eq.~\eqref{Vapprox} as
\begin{eqnarray}
V(\phi_I+\delta\phi)&=&V_0\left[1-\lambda_1\,\delta\phi-\frac{\lambda_5}{5}\delta\phi^5+{\cal O}(\delta\phi^6)\right]\,,
\end{eqnarray}
we can determine the total amount of e-folds $N_{tot.}$ to be
\begin{equation}
N_{tot.}\approx \int_{\phi_0}^{\phi_1}\!d\phi\frac{1}{\sqrt{2\epsilon_V}}\approx \frac{\pi}{\sqrt{2} \sqrt[4]{\lambda_1^3 \lambda_5}}\approx 196\quad.
\end{equation}
This agrees with the numerical results, see Figure~\ref{fig:V8_extra_tilt}. Using this result, we can determine the slow-roll parameters $\epsilon_V$, $\eta_V$ in a similar analytical approximation, and use them to write the spectral index $n_s$ in the limit $N_{tot.}\gg N_{e}$ as
\begin{equation}
n_s\approx 1-\frac{8}{3 N_e}\approx 0.97\,,
\end{equation}
with $N_e\approx 60$. This is in good agreement with current
observations $n_s=0.967\pm0.014$ \cite{Komatsu:2010fb}.

\section{Conclusions}\label{sec:conclusions}
We presented a simple toy model for open inflation after
tunneling. Within a simple polynomial single field potential, the
scalar field tunnels from a false vacuum towards the true vacuum,
subsequently undergoing an extended phase of inflation. While at first
glance, the numerical factors appearing in the potential appear
gigantic, it should be kept in mind that they can be compensated by
rescaling $\phi$ with a number of order $10$ -- apart from the tiny
linear tilt in the potential. However, we only introduced this tilt to
make inflation come to an end. Without it, the field would stay at the
inflection point forever (or until another tunneling event takes it to
the true minimum). Let us stress this point: without introducing the
linear tilt, we would be in a situation where small field inflation
after tunneling suffers from the opposite of an overshoot problem!

We numerically computed the CdL bounce for tunneling from the false
minimum towards the inflection point, and compared it to an analytical
estimate. We then computed the inflationary trajectory numerically and
found that only through introducing a small linear tilt inflation can
come to an end. For the numerical values chosen here, we obtain about
200 efolds of inflation. The scalar spectral index $n_s\approx1$ is in
agreement with current observations. We like to point out that we do
not advertise this model as a realistic candidate ``theory'', but
merely as a toy model which shows that it is quite possible (and
seemingly easy) to obtain a simple model of open small-field inflation
without any overshoot problem after tunneling.

\section*{Acknowledgments}
We are grateful to A. Linde for enlightening comments. This work was supported by the Impuls und Vernetzungsfond of the
Helmholtz Association of German Research Centers under grant
HZ-NG-603, and German Science Foundation (DFG) within the
Collaborative Research Center 676 ``Particles, Strings and the Early
Universe''.

\newpage

\bibliographystyle{JHEP}
\bibliography{single_field_open_inflation}
\end{document}